# TYPE IB/C SUPERNOVAE AND THEIR RELATION TO BINARY STARS[1]

BRUNO LEIBUNDGUT
*European Southern Observatory*
*Karl-Schwarzschild-Strasse 2*
*D-85478 Garching*
*Germany*

**Abstract.**
   The present understanding of type Ib/c supernovae and their connection to interacting binaries is reviewed. The problems of the classification and the lack of well-observed events exclude direct inference of progenitor characteristics. The absence of hydrogen lines in the observed spectrum, nevertheless, requires restricted evolutionary schemes to produce suitable progenitor stars for core collapse explosions with no hydrogen envelope. New relative statistics among the supernova types are presented which indicate that SN Ib/c are on average brighter than SN II, and with the dense sampling of supernova searches in nearby galaxies, a small intrinsic incidence of SN Ib/c is determined. The small rates might be in conflict with the observed ratio of massive stars in binaries in the Galaxy.

## 1. Introduction

Of all supernova classes the type Ib/c supernovae (SN Ib/c) are the most mysterious. Their apparent similarity with SN Ia in light curves and early spectral evolution, while a core collapse probably initiates the explosion, makes them the case of "cross dressing" supernovae. They were discovered as a separate subclass only about a decade ago and the difficulty of distinguishing them clearly from the other classes hamper meaningful statistics.

   The spectrum near maximum light lacks any obvious signs of hydrogen lines and displays remarkably weak lines of Si II ($\lambda$ 6355Å) which made the classification (Harkness & Wheeler 1990) basically one by exclusion of the other supernova types. Confusion arose from the resemblance of the maximum light spectrum of SN Ib/c with the one of SN Ia about four weeks past peak, which led to the expression that SN Ib/c are "born old" (Panagia 1984). The nebular spectrum is distinguished by the strong emission lines of forbidden oxygen and calcium. Again no hydrogen is observed. The optical light curves are almost indistinguishable from the ones of SN Ia (Vacca & Leibundgut 1995), while the near infrared brightness evolution is characteristically different (Elias et al. 1985).

   The discovery of SN 1993J in M81, a type II supernova, has strengthened the connection of SN Ib/c with core collapse events. SN 1993J displayed many signatures of SN Ib/c (Filippenko et al. 1994). The optical light curves are very similar to SN Ib/c (Leibundgut 1994) and the spectrum did develop very strong oxygen and calcium lines in the nebular

---

phase (e.g. Lewis et al. 1994). It has to be emphasized that, despite early expectations, SN 1993J always displayed H$\alpha$ emission (up to 500 days; RGO data archive, Filippenko et al. 1994, Patat et al. 1995).

The progenitors of SN Ib/c have to date been fairly elusive. Several studies have tried to connect SN Ib/c with star formation regions, but so far without conclusive results (Panagia & Laidler 1991, Van Dyk 1992). Recent theoretical work has concentrated on models with stars of small mass which lost all their hydrogen (Nomoto et al. 1990, Woosley et al. 1993, 1994b, Wheeler et al. 1994). To achieve this with current stellar evolution models mass exchange in close binaries is invoked (Nomoto et al. 1994).

Attempts to determine the rates of supernovae have been fairly restricted in the case of SN Ib/c due to the difficulties in separating them from the other classes and the limited sample size (only 32 SN Ib/c have been identified to date). The latter may be caused by the apparent faintness (relative to the other supernovae) and, possibly, large extinction from the environment of the explosions, or an intrinsic rareness of the phenomenon. An attempt to resolve this question will be made.

In the following we will present the current classification scheme and discuss its status within the framework of our understanding of supernovae (§2). Especially the subclassifications into SN Ib and SN Ic will be reviewed critically. Some very simple supernova statistics are presented in section 3 to investigate the nature of SN Ib/c relative to other supernova types. The discussion (§4) describes the minimal knowledge achieved so far and some cautious conclusions.

## 2. The nature of SN Ib/c

Several observables have been exploited to determine the nature of SN Ib/c. Among the techniques are association with star formation regions (Panagia & Laidler 1991, Van Dyk 1992, Van Dyk & Filippenko, these proceedings) starting from the notion that no SN Ib/c has ever been observed in an elliptical galaxy (e.g. van den Bergh & Tammann 1991), spectral modeling (Harkness et al. 1987, Swartz et al. 1993, Jeffery et al. 1991, Wheeler et al. 1994), and radio observations indicating dense circumstellar material around the explosion (Van Dyk et al. 1993). A cursory survey of the published data, discovery announcements, and the CfA data archive provided some coarse statistics on observed H$\alpha$ emission within the slit width (typically 1 to 2 arcseconds) of SN Ib/c. For the 25 supernovae since 1983 we find 14 events with narrow H$\alpha$ emission superposed on the supernova spectrum reported. Considering the inhomogeneity with which these data have been assembled (some from long exposures at late phases others from short integrations of bright supernovae near maximum) the agreement with other determinations (Van Dyk 1992, Van Dyk & Filippenko, these proceedings) is remarkable.

The absence of hydrogen has triggered the association of SN Ib/c with Wolf-Rayet stars (e.g. Wheeler & Levreault 1985) and very massive progenitors as well as stars stripped of the hydrogen envelope induced by binary interaction (Nomoto et al. 1994, van den Heuvel 1994). Progress has been slow due to the rareness of well-observed SN Ib/c. The supernovae leading to the introduction of the new subclass still represent the prime examples. The observations of SN 1985F (Filippenko & Sargent 1986), SN 1983N (Panagia 1985), and SN 1984L (Wheeler & Levreault 1985, Uomoto & Kirshner 1985, Schlegel & Kirshner 1989) have been supplemented only with SN 1987M (Filippenko et al. 1990). The bright SN 1994I in M 51 will change this situation profoundly (Wheeler et al. 1994). Thus, most knowledge has been gathered from statistical evidence and interference with other supernova types. The *hybrid* supernovae SN 1987K (Filippenko 1988) and SN 1993J (Lewis et al. 1994) displayed many characteristics of SN Ib/c at late phases and implied a close relation between SN II and SN Ib/c. Nevertheless, SN 1993J exhibited sustained hydrogen emission and never fully changed its appearance to a SN Ib/c (Lewis et al. 1994, Filippenko et al. 1994).

The separation of the subclasses SN Ib, i.e. helium-rich, and SN Ic, helium-poor, has proven to be rather difficult to implement. The spectroscopic differences are subtle and it has never been shown that there could not exist a continuum rather than a dichotomy in the distribution. Harkness et al. (1987) argued on the basis of the strength of the He I lines in the optical spectrum for a separation, but recently Wheeler et al. (1994) prefer a distinction on the basis of the neutral oxygen line at $\lambda$ 7774Å. This absorption is strong in SN Ic while fairly weak in SN Ib. An open issue is also the appearance of the He I 10830Å line. Its strength is largely undetermined for any supernova. SN 1990W has been reclassified to a SN Ib due to the strong emission in this line (Wheeler et al. 1994), although it did not clearly display strong He I lines in the optical. The reclassification is based on spectral synthesis calculations which indicate enhanced He in the optical spectrum.

Additional distinguishing characteristics between the two subtypes are also rather sparse. The declines of the late time light curves are suspected to be steeper for SN Ic while the smaller rates in SN Ib have been interpreted as due to larger envelope masses in the explosions (Swartz & Wheeler 1991). The sampling of light curves of SN Ib/c, however, is at best still marginal (Vacca & Leibundgut 1995) and firm conclusions are not possible yet.

Hydrogen in SN Ib/c is a controversial issue. The identification of H$\alpha$ in the early time spectra of SN Ic (Jeffery et al. 1991, Filippenko 1992) has not been generally accepted (Swartz et al. 1993). The main worry is the expansion velocity measured for hydrogen which appears smaller than for calcium and oxygen contrary to any reasonable explosion model. Swartz et al. (1993) experimented with models including a small amount of hydrogen and found strong inconsistencies with the observed spectra of SN 1987M. SN 1993J with its close resemblance in certain aspects with SN Ib/c but its obvious hydrogen emission has further complicated the interpretation.

Another possible distinction pattern in the peak light spectra has now been proposed by Wheeler et al. (1994). The lines of the C II ($\lambda$ 6580Å) and Si II ($\lambda$ 6355Å) appear in SN Ic like SN 1994I and SN 1987M, while an absorption line observed at almost the same position in the spectrum of SN Ib is interpreted to arise from H$\alpha$, e.g. SN 1983N (Wheeler et al. 1994). If these identifications are correct, the evolution of these lines is expected to be different and might provide a rather stringent test on the nature of the two subclasses.

Classification of supernovae is a difficult issue. For analyses connecting supernova explosions with global parameters, like star formation rates or chemical enrichment, a simple scheme which can quickly and easily separate clearly distinct objects is needed. For a detailed understanding of the physics of the individual objects a more specific description is called for. The difficulty in separating type Ib objects from type Ic probably outlines this border. Spectral synthesis calculations are needed to find the subtle distinctions between the two classes. This precludes classifications with spectra of newly discovered supernovae at the telescope, as it has been done in most cases in the past. Statistics based on supernova catalogs become difficult with such disparate classification systems.

## 3. Statistics of SN Ib/c

Two main problems plague supernova statistics. First, the numbers are rather small and meaningful statistics are hard to find. This is especially so when further subdivision of the samples is attempted, a procedure obviously necessary to investigate the different underlying physics of the events. Second, the strong selection biases inherent in the detection mechanisms and the classification procedures as described for the case of SN Ib/c above. The basic ingredients for the statistics are (see Strom 1994 for an excellent review): the observed supernova rate, i.e. the number of supernovae per unit time interval, and the corrections depending on the form of the luminosity profiles and distribution of the individual objects. The tricky part is how to correct to find the true number of explosions in a galaxy (e.g. Cappellaro et al. 1993, van den Bergh & McClure 1994).

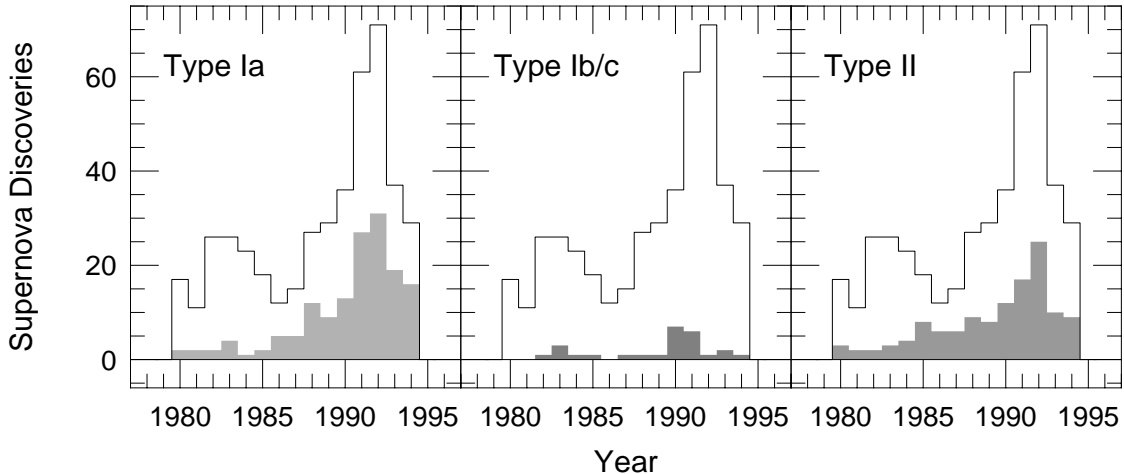

*Figure 1.* Supernova discoveries between 1980 and 1994. The data and classifications are from the Asiago Supernova Catalog (Barbon et al. 1989) for all supernovae before 1989 and according to IAU Circulars and the literature thereafter. The line histogram shows the total number of supernovae discovered per calendar year and the shaded areas the supernovae of the particular type.

With the limited number of SN Ib/c known to date (32 objects in total) it is difficult to derive significant results from statistics. Since in the past the lack of the fine tool of spectral synthesis prevented classification of the objects into SN Ib and SN Ic, we will apply statistics only to the overall sample of known SN Ib/c. An important deficit of this study will also be objects misclassified as SN Ia, due to their similar spectra and light curves throughout the peak phase, and the separation from SN II has become difficult as there are cases where a clear association to either class is not obvious. The examples of SN 1987K and SN 1993J should serve as warnings. Unless a sample can be constructed which has a uniform classification applied to it, we will have to deal with a rather inhomogeneous data set. We have chosen to use the classification near maximum light reported in the literature for the selection of our samples.

An overall picture of the supernova discovery rates and the number of the individual classes for the last 15 years is presented in Figure 1. It is evident how limited the number of SN Ib/c is relative to the other two classes. There is also an interesting trend discernible. While SN Ia and SN II were found roughly at a constant fraction of the total number of discoveries, the number of SN Ib/c appears anti-cyclic at the beginning of this decade. The distributions in Fig. 1 are mainly the signatures of two complementary supernova searches. The Calan-Tololo supernova search (Hamuy et al. 1993) was targeted at rather faint, distant supernovae while the Berkeley supernova search (Perlmutter et al. 1990) was aimed to find supernovae in relatively nearby galaxies. The former was running during the years 1990-1993 (Hamuy et al. 1995) and has been a major contributor to the pronounced peak of discoveries. The latter, employing an unfiltered CCD, started seriously 1988 and ended 1991. The Berkeley search is also responsible for the large number of SN Ib/c at the turn of the decade (Muller et al. 1992). The red sensitivity of CCDs has been proposed as the cause for the increased *relative* SN Ib/c rate found by the Berkeley group (Cappellaro et al. 1993), but only six out of the 15 SN Ib/c between 1988 and 1991 are from this search. No real explanation is offered here for the strange distribution of SN Ib/c discoveries.

Relative changes among the various supernova classes can also be garnered from Fig. 1.

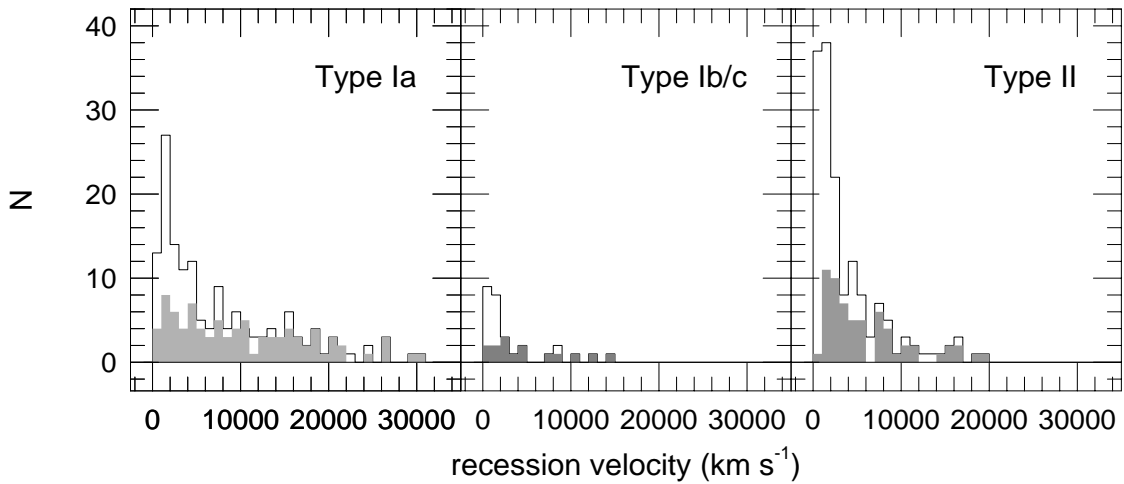

*Figure 2.* Histograms of the distribution of supernovae in distance bins. The shaded areas are statistics starting only 1989. The bin width is 1000 km s$^{-1}$. Thus only the two closest bins are likely to be sensitive to peculiar galaxy motions.

The increased rate of SN Ia is most likely due to their intrinsic brightness and the fainter detection limits of recent searches. The increase of SN II discoveries is more moderate while the SN Ib/c have fluctuated substantially. The lower discovery rate of SN Ib/c compared with SN II are the signature of the combination of two effects. For illustration consider two extreme cases. Let's first assume SN Ib/c have the same luminosity as SN II which means that the former must be rarer as the detection probability is roughly the same. On the other hand, if both types have about equal rates, Fig. 1 implies SN Ib/c to be fainter (either because they are less luminous or suffer from significant extinction). Complications arise from the differently shaped light curves. If SN II remain brighter for an extended period, which is certainly the case for the plateau-type light curves, their discovery chance is increased proportionally.

To investigate the relative differences among supernova classes further we employ a very simple statistical scheme. The only inputs required are a classification of the supernova and the recession velocity of the parent galaxy. The distribution for each type is presented in Figure 2. We have chosen to display all samples on the same scale to emphasize the variation in the *total* number of objects known in each class. The small sample of known SN Ib/c is once more striking. This is also true for the more recent samples starting in 1989 (the shaded areas). In passing we note that the last five years contain 44 percent of all classified supernovae which makes this subsample statistically useful, especially when considerations like the restricted number of classifying people are taken into account. Not shown are the few objects with redshifts substantially larger than 0.1. There are several identifications of objects at higher redshift with rather uncertain classifications. They include a few SN Ia (SN 1988U, and SN 1994F; possibly SN 1992bi, SN 1994G and SN 1994H), one SN Ib/c (SN 1992ar), and one probable SN II (SN 1988T).

Not surprisingly all distributions in Fig. 2 are peaked towards small distances. In particular the overall samples offer very little room for distinction among the distributions. The situation is slightly different if one considers only the more recent supernovae. With the increase of the limiting magnitudes of the searches over the past few years, the distribution of SN Ia is much flatter than before. In other words, most of the more distant supernovae have been found in recent years. Although this equally applies to all classes the effect is less pronounced for SN II, which still are discovered preferentially within a distance of $cz \leq 10000$ km s$^{-1}$. The higher peak luminosity of SN Ia is easily inferred from this diagram. The histogram for SN Ib/c is clearly less peaked than the one of SN II, although the small numbers make this comparison rather shaky. What is further complicating the picture is the discovery of SN 1992ar (Hamuy et al. 1992). This SN Ib/c was located in a galaxy with a redshift of 0.145 and appears to have been similar in brightness to SN Ia. The picture presented in Fig. 2 is then not so simple anymore. But the fact of a SN Ib/c with such a high luminosity hints to the possibility that these objects are in their mean not fainter than SN II.

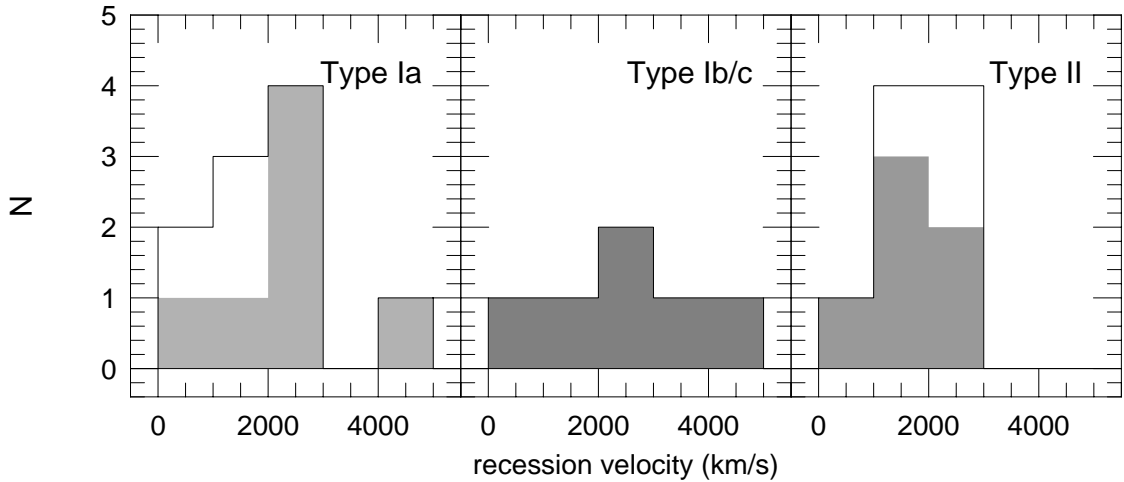

*Figure 3.* Same as Fig. 2 for the supernovae of Muller et al. (1992), shaded area, and the sample including the supernovae discovered during the first year of LOSS (Treffers et al. 1994).

A high *relative* rate for SN Ib/c was found in data extracted from the Berkeley Supernova Search (Muller et al. 1992). Although plagued with very small numbers (only a total of 12 objects were included in their analysis), the highest total number of any subclass was found for SN Ib/c. While there is a fair agreement for the rates of SN Ia and SN II Muller et al. find an enhancement in their SN Ib/c rate by a factor of three over other determinations (van den Bergh & Tammann 1991, Cappellaro et al. 1993, van den Bergh & McClure 1994). We tried to assess this result with the statistics presented above. Figure 3 displays the number of supernovae in the different distance bins. For these diagrams we have combined the samples from Muller et al. and the supernovae discovered during the first year of the Leuschner Observatory Supernova Search (LOSS; Treffers et al. 1994). The two searches have employed the same equipment but slightly different detection algorithms. LOSS is not fully automatic but rather provides a list of candidates every morning which are then inspected by an observer. While the original Berkeley search was performed with the unfiltered CCD, LOSS is using an R filter. The limiting magnitudes of the two searches are probably comparable at $R = 17$. An important aspect of these searches are the high frequency with which the galaxies are observed (typically once a week) which for a distant limited sample of galaxies (like the one employed in LOSS) reduces the dependence on light curve shapes considerably.

While the numbers are still very small we find that searches targeted at nearby galaxies do indeed find a fair fraction of SN Ib/c in the total sample. The distribution in Fig. 3 is 10:6:9 (SN Ia:SN Ib/c:SN II). Also the clustering towards small distances is obvious for SN Ia and SN II. Not so clear is the distribution for SN Ib/c which appears rather flat. Incidentally, LOSS has only added SN Ia and SN II to the total sample during the first year, but no SN Ib/c. Most likely this is due to small number statistics. The distance distribution clearly indicates that SN Ib/c certainly are *not* fainter than SN II. The argument that the rate of SN Ib/c should be higher than the rates for the other supernova classes can not be supported. Fig. 3 clearly demonstrates that the argument of Muller et al. (1992) does not hold for their own data. We then conclude that the rate of SN Ib/c is certainly smaller

TABLE 1. Galaxies with multiple supernova events

| SN type | $N_{SNe}$ | percentage | SN type | $N_{SNe}$ | percentage |
| --- | --- | --- | --- | --- | --- |
| I | 6 | 17 | Ia | 16 | 20 |
| II | 20 | 57 | Ib/c | 3 | 4 |
| unclassified | 8 | 23 | I | 7 | 9 |
| peculiar | 1 | 3 | II | 21 | 26 |
|  |  |  | unclassified | 35 | 41 |

Note: 2 galaxies with 6 supernovae
2 galaxies with 4 supernovae
5 galaxies with 3 supernovae

Note: 41 galaxies with 2 supernovae

than for SN II and also that they have a comparable detection chance to SN II.

A last test to compare the relative frequencies of supernovae is the comparison of rates in galaxies with multiple supernova events (Richter & Rosa 1988). This assumes that the survey time for all parts of a galaxy are equal and that global effects triggering supernova events, e.g. starbursts, are not influencing the relative rates between the different kind of core collapse supernovae. There have now been a fair number of galaxies with multiple supernova events over the last century. The distributions by type are listed in Table 1. For the 9 galaxies with more than two supernovae (a total of 35 events) SN II are three times as frequent as SN I. We have not distinguished among SN Ia and SN Ib/c as most events are of historical character. Unless a large fraction of SN II are misclassified SN I, there is a strong preponderance of hydrogen-displaying supernovae. Inclusion of the 41 galaxies with 2 events increases the fraction of unclassified supernovae, but still shows that SN II have been discovered at least twice more often than SN Ib/c.

Overall, it appears clear that SN Ib/c are observed less frequent than SN II. This is implied by the smaller total number of objects while the mean distance in the samples appears to be larger (Figs. 2 and 3). With similar detection probabilities (especially in densely sampled searches like the two Leuschner projects) we have to conclude that SN Ib/c are intrinsically rarer than SN II (and SN Ia).

## 4. Discussion

To determine the stellar progenitors of supernovae has proven a very difficult enterprise. The explosion destroys the progenitor and the chemistry changes erase many traces of the composition of the progenitor star. This is much more the case for explosions inside small envelopes, typically assumed for SN Ia and SN Ib/c. Thus, it is very difficult to estimate the characteristics of the star before it hit catastrophe. Observing supernova progenitors before the explosion is restricted to the nearest galaxies. The progenitor of SN 1987A is the only clear identification to date, while a progenitor of the type Ic SN 1994I in M51 could not be identified (Kirshner et al. 1994). A possible identification has been reported for the progenitor of SN 1993J (Aldering et al. 1994). Photometry of the stellar object coincident with the position of SN 1993J indicates an excess of blue light compared with the progenitor structure inferred from models of the supernova (e.g. Wheeler & Filippenko 1994, Woosley et al. 1994a). The blue excess could be due to a nearby cluster of blue stars or a massive, hot companion. Although single star models can be devised to retain only a small hydrogen envelope such an evolution occurs more naturally in a binary star scenario (Woosley et al. 1994b, Nomoto et al. 1993, 1994). The observed non-spherical distribution of matter in the ejecta of SN 1993J (Fransson et al. 1994) is also indicative of a binary system. This is to illustrate that the borders of supernova classes not necessarily outline the different evolutionary channels. Nevertheless, we might derive very general conclusions, if we assume that SN Ib/c are most likely originating in binary systems (Nomoto et al. 1994, Wheeler et al. 1994, Woosley et al. 1994a,b). Paradoxically, this statement very strongly relies on the identification of one SN II (SN 1993J) with a binary progenitor.

The current study restricted itself to find the relative differences in the occurrence of core collapse supernovae. Taking the results of section 3, we have to conclude that SN Ib/c are on average brighter at maximum than SN II as they are observed at larger mean distances. We cannot directly infer that SN Ib/c are rarer than SN II due to the distinct light curve shapes of the two classes. Since SN Ib/c drop about two magnitudes in 20 days in blue light whereas many SN II remain on a high-luminosity plateau for up to 2 months (Patat et al. 1994) the chance of discovery is strongly enhanced for SN II. At redder wavelengths, as is the case of the Berkeley searches, this effect is less pronounced due to the redder colors of SN Ib/c at maximum. Thus, we believe that with SN Ib/c on average brighter than SN II the number statistics reflect a truly smaller incidence of SN Ib/c compared to SN II. Such a small frequency has been found in several other studies before (van den Bergh & McClure 1994, Cappellaro et al. 1993) contrasting with the rates derived by Muller et al. (1992). An interesting unresolved question remains why SN Ib/c discoveries peaked in a different year than the discoveries of all other types (Fig. 1). It might be that the absolute magnitudes at peak and the rareness of the events resulted in a strong bias against these objects in the nearby and distant searches conducted at the beginning of the decade.

The rareness of SN Ib/c and the tentative identification of these events with progenitors in interacting binary systems means that this evolutionary channel is constraint fairly strongly and very specific scenarios are needed (e.g. van den Heuvel 1994). Most models favor evolutions involving common envelope phases to strip stars from their H and He envelopes. With low rates for SN Ib/c an apparent conflict is further obtained in the estimate of the progenitor systems and the observed ratio of SN Ib/c to SN II. Using the formalism described in van den Heuvel (1994) we find that only roughly a quarter of all stars with M>10 $M_\odot$ are in interacting binary systems. Some explosions in close binary systems might, however, retain enough hydrogen to disguise as SN II, as observed in SN 1993J.

Studying SN rates to learn about local properties like the fraction of close binaries assumes that what we find in a morphological mixture of external galaxies still is applicable to the Galaxy. SN rates are derived from an ensemble of very distinct properties and it is by no means obvious that an average over all possible solutions provides something close to the situation in the Galaxy. Lacking other more direct observational routes to determine the fate of stellar evolution we have to make a big swing through the nearby universe and stretch our imagination to obtain any results.